\documentclass[aps,twocolumn,showpacs,preprintnumbers]{revtex4}

\usepackage{braket} 
\usepackage{lipsum}
\usepackage{graphicx}  
\usepackage{subfigure}
\usepackage{multirow}
\usepackage{float}
\usepackage{fancyhdr}
\usepackage{longtable}
\usepackage{parskip}
\usepackage[T1]{fontenc}
\usepackage{dcolumn}   
\usepackage{booktabs}
\usepackage{bm}        
\usepackage{amsfonts}  
\usepackage{amsmath}   
\usepackage{amssymb}   

\newcommand{\pwisein}{\left\{ \begin{array}{ll}}
\newcommand{\pwiseout}{\end{array}\right.}

\begin{document}

\title{The $Z_b$ states as the mixture of the molecular and diquark-anti-diquark components within the effective field theory}

\author{Wei He$^{1,2}$}
\email[]{hewei1999@outlook.com}
\author{De-Shun Zhang$^{1,2}$}
\email[]{220220940071@lzu.edu.cn}
\author{Zhi-Feng Sun$^{1,2,3,4}$}
\email[Corresponding Author: ]{sunzf@lzu.edu.cn}

\affiliation {\it$^1$School of Physical Science and Technology, Lanzhou University, Lanzhou 730000, China\\
$^2$Research Center for Hadron and CSR Physics, Lanzhou University and Institute of Modern Physics of CAS, Lanzhou 730000, China\\
$^3$Lanzhou Center for Theoretical Physics, Key Laboratory of Theoretical Physics of Gansu Province, and Key Laboratory of Quantum Theory and Applications of the Ministry of Education, Lanzhou University, Lanzhou, 730000, China\\
$^4$Frontiers Science Center for Rare Isotopes, Lanzhou University, Lanzhou, Gansu 730000, China
}

\date{\today}

\begin{abstract}  
In this study, we reconsider the states $Z_b(10610)$ and $Z_b(10650)$ by investigating the presence of diquark-anti-diquark components as well as the hadronic molecule components in the framework of effective field theory. The different masses of pseudoscalar mesons such as $\pi^{0}$, $\eta_{8}$, and $\eta_{0}$, as well as vector mesons like $\rho^{0}$ and $\omega$ violate the OZI rule that is well depicted under the $[U(3)_L\otimes U(3)_R]_{global}\otimes [U(3)_V]_{local}$ symmetry. To account for the contribution of intermediate bosons of heavy masses within the OBE model, we introduce an exponential form factor instead of the commonly used monopole form factor in the past. By solving the coupled-channel Schr\"{o}dinger equation with the Gaussian expansion method, our numerical results indicate that the $Z_b(10610)$ and $Z_b(10650)$ states can be explained as hadronic molecules slightly mixing with diquark-anti-diquark states.   
\end{abstract}

\maketitle 

\section{Introduction}
In the past decades, a series of quarkonium-like states were discovered. In the $b\bar{b}$ sector, Belle Collaboration reported two charged bottomonium-like states which are known as $Z_b(10610)$ and $Z_b(10650)$ \cite{Belle:2011aa} in 2011. Both of them were observed in $\Upsilon(5S)\to \pi^{\pm}h_{b}(mP) \ (m=1,2)$ and $\Upsilon(5S)\to\pi^{\pm}\Upsilon(nS) \ (n=1,2,3)$, respectively. Later, Belle confirmed their observations \cite{Belle:2014vzn,Belle:2015upu}. The next year of the first discovery, the neutral state $Z_b^0(10610)$ was found in the $\Upsilon(5S)\to \Upsilon(2S,3S)\pi^0\pi^0$ decay \cite{Belle:2012glq}. The masses and widths of these states listed in PDG (Particle Data Group) are shown below
{
\footnotesize
\begin{align}
    M_{Z_b^\pm}&=10607.2\pm2.0\ \mathrm {MeV}, \quad\Gamma_{Z_b^\pm}=18.4\pm2.4\ \mathrm{MeV},\nonumber\\
    M_{Z_b^0}&=10609\pm 4.0\pm 4\ \mathrm {MeV},\nonumber\\
    M_{Z_b^{\prime \pm}}&=10652.2\pm1.5\ \mathrm {MeV}, \quad\Gamma_{Z_b^{\prime \pm}}=11.5\pm2.2\ \mathrm{MeV} \nonumber
\end{align}} 
with the quantum numbers $I^G(J^P)=1^+(1^+)$. For simplicity, here we label the two states $Z_b(10610)$ and $Z_b(10650)$ by $Z_b$ and $Z_b^\prime$, respectively.

Theoretical research had already been performed before the observations of the $Z_b$ states. The authors in Ref. \cite{Liu:2008fh,Liu:2009qhy} indicated that there may exist a loosely bound S-wave $B\bar{B}^{*}/B^{*}\bar{B}$ molecular state. 

After the observation, the explanations of the nature of the $Z_b$ states were proposed through different assumptions and theoretical methods. Since the masses of the $Z_b(10610)$ and the $Z_b(10650)$ are close to the $B\Bar{B}^*$ and $B^*\Bar{B}^*$ thresholds, they are good candidates of $B\Bar{B}^*$ and $B^*\Bar{B}^*$ molecular states \cite{Bondar:2011ev,Cleven:2011gp,Mehen:2011yh,Nieves:2011zz,Ohkoda:2011vj,Sun:2011uh,Voloshin:2011qa,Yang:2011rp,Zhang:2011jja,Dong:2012hc,Li:2012wf,Li:2012uc,Ohkoda:2012rj,Cui:2011fj,Cleven:2013sq,Cleven:2013rkf,Ohkoda:2013cea,Mehen:2013mva,Wang:2013daa,Dias:2014pva,Wang:2014gwa,Chen:2015ata,Kang:2016ezb,Goerke:2017svb,Gutsche:2017oro,Ozdem:2017exj,Voloshin:2017gnc,Yang:2017rmm,Wang:2018pwi,Wang:2018jlv,Guo:2013sya,Baru:2020ywb,Zhang:2022hfa,Wu:2022hck,Cheng:2023vyv}. However, tetraquark interpretations including diquark-anti-diquark explanation can not be ruled out \cite{Ali:2011ug,Esposito:2014rxa,Maiani:2017kyi,Cui:2011fj,Agaev:2017lmc,Wang:2013zra,Ali:2014dva,Ke:2012gm,Ozdem:2017exj}. As a consequence, in this work we study these two states in the picture of the mixture of molecular and diquark-anti-diquark components, which is used to investigate the nature of the $Z_{cs}$ states observed by LHCb in our previous work \cite{Cao:2022rjp}. 

The hadronic molecule has been proposed based on the study of deuteron composed of a proton and a neutron. And this kind of topic has been widely discussed \cite{Chen:2016qju,Chen:2022asf,Dong:2021bvy,Meng:2022ozq} within different methods, especially after the observation of $X(3872)$ in 2003 \cite{Belle:2003nnu}. On the other hand, the concept of the diquark-anti-diquark state was proposed for the first time by Maiani $et\ al.$ \cite{Maiani:2004uc,tHooft:2008rus} following the revitalization of interest on the $\sigma$ meson. In this work, we use both the molecular state and the diquark-anti-diquark state, and make the calculation in the framework of the effective field theory. In this way, the mesons and the diquarks are viewed as point-like particles, which finally form the color singlet system. The forces between these clusters are provided by exchanging pseudoscalar and vector mesons as well as scalar and axial-vector diquarks. 

In order to calculate the effective potentials in the coordinate space, form factors for each vertex are needed, such that the high-momentum contributions are suppressed. In the works related to the One-Boson-Exchange model in the past, the monopole form factor is introduced (see the review \cite{Chen:2016qju}). However, in some of our cases, since the exchanged particles' masses are not so small, the monopole form factor does not work very well for it suppressing the corresponding potentials by its numerator. So we introduce the exponentially parameterized form factor and obtain the analytical expressions of the potential in the coordinate space. Considering both the S- and D-wave contributions, we solve the coupled channel Schr\"{o}dinger equation to see the existence of the composite particles.

The structure of this paper is organized as follows. After the introduction, the theoretical framework is presented in Sec. \ref{formalism}. The results and discussion are shown in Sec. \ref{results}. Finally, a brief summary is given in Sec. \ref{summary}.

\section{Formalism}\label{formalism}
\subsection{Wave functions}
We give here the flavour wave functions of the negative and neutral $B\bar{B}^{*}/B^{*}\bar{B}$ and $B^{*}\Bar{B}^{*}$ systems constructed in Ref. \cite{Sun:2011uh}
\begin{equation}\footnotesize
\begin{aligned}
    \ket{Z_{B\bar{B}^{*}/B^{*}\bar{B}}^{-}}=&\frac{1}{\sqrt{2}}(\ket{B^{*-}B^{0}}+c\ket{B^{-}B^{*0}}),\\
    \ket{Z_{B\bar{B}^{*}/B^{*}\bar{B}}^{0}}=&\frac{1}{2}[\ket{B^{*+}B^{-}}-\ket{B^{*0}\Bar{B}^{0}}+c(\ket{B^{+}B^{*-}}\\
    &-\ket{B^{0}\Bar{B}^{*0}})],\\
    \ket{Z_{B^{*}\bar{B}^{*}}^{-}}=&\ket{B^{*-}B^{*0}},\\
    \ket{Z_{B^{*}\bar{B}^{*}}^{0}}=&\frac{1}{\sqrt{2}}(\ket{B^{*+}B^{*-}}-\ket{B^{*0}\Bar{B^{*0}}}).
\end{aligned}
\end{equation}
The flavour wave functions of diquark and anti-diquark systems are constructed in analogy to the meson-meson systems by swapping $b$ quark and $\Bar{b}$ quark,
\begin{equation}\footnotesize
\begin{aligned}
     \ket{Z_{S\bar{A}/A\Bar{S}}^{-}}&=\frac{1}{\sqrt{2}}(\ket{\bar{A}_{bu}S_{bd}}+c\ket{\bar{S}_{bu}A_{bd}}),\\
     \ket{Z_{S\bar{A}/A\Bar{S}}^{0}}&=\frac{1}{2}[\ket{A_{bu}\bar{S}_{bu}}-\ket{A_{bd}\bar{S}_{bd}}+c(\ket{S_{bu}\bar{A}_{bu}}-\ket{S_{bd}\bar{A}_{bd}})],\\
    \ket{Z_{A\bar{A}}^{-}}&=\ket{A_{bd}\bar{A}_{bu}},\\
    \ket{Z_{A\bar{A}}^{0}}&=\frac{1}{\sqrt{2}}(\ket{A_{bu}\bar{A}_{bu}}-\ket{A_{bd}\bar{A}_{bd}}).  
\end{aligned}
\end{equation}
The value of $c$ depends on the G-parity, i.e.,  $c=\pm 1$ corresponds to $G=\pm 1$. Here we only pay attention to the situation of $c=+1$, since the G-parity of the considered systems are all $+1$.

In this work, both S- and D-wave interactions between the composed particles are considered. In general, the $Z_b$ and $Z_b^\prime$ states can be expressed as 
{\small
\begin{eqnarray}
|Z_b\rangle &=&\left(
\begin{array}{c}
Z_{B\Bar{B}^*/B^*\Bar{B}}^-(^3S_1)\\
Z_{B\Bar{B}^*/B^*\Bar{B}}^-(^3D_1)\\
Z_{B^*\Bar{B}^*}^-(^3S_1)\\
Z_{B^*\Bar{B}^*}^-(^3D_1)\\
Z_{S\Bar{A}/A\Bar{S}}^-(^3S_1)\\
Z_{S\Bar{A}/A\Bar{S}}^-(^3D_1)\\
Z_{A\Bar{A}}^-(^3S_1)\\
Z_{A\Bar{A}}^-(^3D_1)
\end{array}
\right),\
|Z_b^\prime\rangle =\left(
\begin{array}{c}
Z_{B^*\Bar{B}^*}^-(^3S_1)\\
Z_{B^*\Bar{B}^*}^-(^3D_1)\\
Z_{S\Bar{A}/A\Bar{S}}^-(^3S_1)\\
Z_{S\Bar{A}/A\Bar{S}}^-(^3D_1)\\
Z_{A\Bar{A}}^-(^3S_1)\\
Z_{A\Bar{A}}^-(^3D_1)
\end{array}
\right).\nonumber\\
\end{eqnarray}
}
Note that for $Z_b^\prime$, $|^5D_1\rangle$ state is forbidden due to its G-even parity.

\subsection{ The Effective Lagrangians and Coupling Constants}
Next we introduce the interactions of meson-meson and diquark-anti-diquark by constructing the corresponding Lagrangians. 

As we all know, each quark is a color triplet resulting in a diquark being a color antitriplet or sextet. The interaction between the two quarks of an antitriplet is attractive, while that of a sextet is repulsive. Consequently, we only consider the effective Lagrangian containing antitriplet. The diquark fields are depicted as 
\begin{eqnarray}
S^a&=&\left(\begin{array}{ccc}
     0&S_{ud}&S_{us}\\
     -S_{ud}&0&S_{ds}\\
     -S_{us}&-S_{ds}&0
    \end{array}\right)^a,\\
A_\mu^a&=&\left(\begin{array}{ccc}
     A_{uu}&\frac{1}{\sqrt{2}}A_{ud}&\frac{1}{\sqrt{2}}A_{us}\\
     \frac{1}{\sqrt{2}}A_{ud}&A_{dd}&\frac{1}{\sqrt{2}}A_{ds}\\
     \frac{1}{\sqrt{2}}A_{us}&\frac{1}{\sqrt{2}}A_{ds}&A_{ss}
    \end{array}\right)_\mu^a,\\
S_b^a&=&\left(\begin{array}{ccc}
     S_{bu}&S_{bd}&S_{bs}
    \end{array}\right)^a,\\
A_{b\mu}^a&=&\left(\begin{array}{ccc}
     A_{bu}&A_{bd}&A_{bs}
    \end{array}\right)_\mu^a,
\end{eqnarray}
where $S^a$ is the light scalar diquark, $A_{\mu}^a$ the light axial vector diquark, $S_b$ the bottomed scalar diquark, and $A_{b\mu}^a$ the bottomed axial vector diquark. The superscript $a=1,2,3$ is the color index. The meson fields read
{\footnotesize
\begin{eqnarray}
\Phi&=&\left(\begin{array}{ccc}
\frac{\sqrt{3}\pi^0+\eta_8+\sqrt{2}\eta_0}{\sqrt{3}}&\sqrt{2}\pi^+ &\sqrt{2}K^{+}\\
\sqrt{2}\pi^-&\frac{-\sqrt{3}\pi^0+\eta_8+\sqrt{2}\eta_0}{\sqrt{3}}&\sqrt{2}K^{0}\\
\sqrt{2}K^{-}&\sqrt{2}\bar{K}^{0}&\frac{-2\eta_8+\sqrt{2}\eta_0}{\sqrt{3}}
\end{array}\right),\label{eq7} \\
V_\mu&=&\frac{g_V}{\sqrt{2}}\left(\begin{array}{ccc}
\frac{1}{\sqrt{2}}(\rho^0+\omega)&\rho^+ &K^{*+}\\
\rho^-&-\frac{1}{\sqrt{2}}(\rho^0-\omega)&K^{*0}\\
K^{*-}&\bar{K}^{*0}&\phi
\end{array}\right)_\mu,\\
P&=&(B^{-},\bar{B}^0,\bar{B}_{s}^{0}),\\
P^{*}_{\tau}&=&(B^{*-},\bar{B}^{*0},\bar{B}_{s}^{*0})_{\tau}
\end{eqnarray}
}
with $\Phi$ and $V_\mu$ the light pseudoscalar and vector, $P$ and $P^*_\tau$ the bottomed pseudoscalar and vector.

Considering the $[U(3)_L\otimes U(3)_R]_{global}\otimes [U(3)]_{local}$ symmetry \cite{Bando:1985rf,Harada:2003jx}, parity and charge conjugation, the Lagrangians containing mesons and diquarks are shown as follows
\begin{eqnarray}
\mathcal{L}_{1}&=&-\frac{\beta}{m_P}(iP\hat{\alpha}_{\|\mu}D^{\mu}P^{\dag}+h.c.)-2g(iP\hat{\alpha}_{\perp\mu}P^{*\mu\dag}\nonumber\\
&&+h.c.)-\frac{g}{m_{P^{*}}}(\xi^{\mu\nu\alpha\beta}P^{*}_{\nu}\hat{\alpha}_{\perp\alpha}D_{\mu}P^{*\dag}_{\beta}+h.c.)\nonumber\\
&&+\frac{\beta}{m_{P^*}}iP^{*}_{\nu}\hat{\alpha}_{\|}^{\mu}D_{\mu}P^{*\nu\dag}+h.c.),\\ 
\mathcal{L}_{2}&=&e_1(iPD_\mu S^aA_b^{a\mu\dag}-iA_b^{a\mu} D_\mu S^{a\dag} P^\dag)\nonumber\\
&&+e_2(iPA^a_\mu D^\mu S_b^{a\dag}-iD^\mu S^a_bA^{a\dag}_\mu P^\dag)\nonumber\\
&&+e_3(\epsilon^{\mu\nu\alpha\beta}PA^a_{\mu\nu}A_{b\alpha\beta}^{a\dag}+\epsilon^{\mu\nu\alpha\beta}A^a_{b\alpha\beta}A^{a\dag}_{\mu\nu}P^\dag)\nonumber\\
&&+e_4(iP^*_\mu D^\mu S^aS_b^{a\dag}-iS^a_bD^\mu S^{a\dag} P^{*\dag}_\mu)\nonumber\\
&&+e_5(\epsilon^{\mu\nu\alpha\beta}P_\mu^*D_\nu S^aA_{b\alpha\beta}^{a\dag}+\epsilon^{\mu\nu\alpha\beta}A^a_{b\alpha\beta}D_\nu S^{a\dag} P_\mu^{*\dag})\nonumber\\
&&+e_6(\epsilon^{\mu\nu\alpha\beta}P_\mu^*A^a_{\nu\alpha}D_\beta S_b^{a\dag}+\epsilon^{\mu\nu\alpha\beta}D_\beta S^a_bA_{\nu\alpha}^{a\dag} P_\mu^{*\dag})\nonumber\\
&&+e_7(iP^*_\mu A^{a\mu\nu}A_{b\nu}^{a\dag}-iA^a_{b\nu}A^{a\mu\nu\dag}P^{*\dagger}_\mu)\nonumber\\
&&+e_8(iP^*_\mu A^a_\nu A_b^{a\mu\nu\dag}-iA_b^{a\mu\nu}A_\nu^{a\dag} P^{*\dag}_\mu)\nonumber\\
&&+e_9(iP^*_{\mu\nu}A^{a\mu} A_b^{a\nu\dag}-iA_b^{a\nu} A^{a\mu\dag}P^{*\dag}_{\mu\nu}),\label{eqLagrangian2}\\
\mathcal{L}_{3}&=&h_{1}(iS^a_{b}\hat{\alpha}^{\mu T}_{\|}D_{\mu}S_{b}^{a\dagger}-iD_{\mu}S^a_{b}\hat{\alpha}^{\mu T}_{\|}S_{b}^{a\dagger})\nonumber\\
&&+h_{2}(\epsilon^{\mu\nu\alpha\beta}A^a_{b\mu\nu}\hat{\alpha}^{\mu T}_{\|\alpha}D_{\beta}S_{b}^{a\dagger}+\epsilon^{\mu\nu\alpha\beta}D_{\beta}S^a_{b}\hat{\alpha}^{\mu T}_{\|\alpha}A_{b\mu\nu}^{a\dagger})\nonumber\\
&&+h_{3}(iA^a_{b\mu}\hat{\alpha}^{\mu T}_{\bot}S_{b}^{a\dagger}-iS^a_{b}\hat{\alpha}^{\mu T}_{\bot}A_{b\mu}^{a\dagger})\nonumber\\
&&+h_{4}(iA^a_{b\mu}\hat{\alpha}^{T}_{\|\nu}A_{b}^{a\mu\nu\dagger}-iA_{b}^{a\mu\nu}\hat{\alpha}^{T}_{\|\nu}A_{b\mu}^{a\dagger})\nonumber\\
&&+h_{5}(\epsilon^{\mu\nu\alpha\beta}A^a_{b\mu}\hat{\alpha}^{T}_{\bot\nu}A_{b\alpha\beta}^{a\dagger}\nonumber\\
&&+\epsilon^{\mu\nu\alpha\beta}A^a_{b\alpha\beta}\hat{\alpha}^{T}_{\bot\nu}A_{b\alpha\mu}^{a\dagger}),\label{eqLagrangian3}
\end{eqnarray}
where
\begin{eqnarray}
D_\mu P&=&\partial_\mu P+iP\alpha^\dag _{\| \mu}=\partial_\mu P+iP\alpha_{\| \mu},\\
D_\mu P^*_\tau&=&\partial_\mu P^*_\tau+iP^*_\tau\alpha^\dag _{\| \mu}=\partial_\mu P^*_\tau+iP^*_\tau\alpha_{\| \mu},\\
\alpha_{\bot\mu}&=&(\partial_\mu \xi_R\xi_R^\dag-\partial_\mu \xi_L\xi_L^\dag)/(2i),\\
\alpha_{\|\mu}&=&(\partial_\mu \xi_R\xi_R^\dag+\partial_\mu \xi_L\xi_L^\dag)/(2i),\\
\hat{\alpha}_{\bot\mu}&=&(D_\mu \xi_R\xi_R^\dag-D_\mu \xi_L\xi_L^\dag)/(2i),\\
\hat{\alpha}_{\|\mu}&=&(D_\mu \xi_R\xi_R^\dag+D_\mu \xi_L\xi_L^\dag)/(2i),\\
\xi_L&=&e^{i\sigma/F_\sigma}e^{-i\Phi/(2F_\pi)},\label{eq21}\\
\xi_R&=&e^{i\sigma/F_\sigma}e^{i\Phi/(2F_\pi)},\label{eq22}\\
A^a_{\mu\nu}&=&D_\mu A^a_\nu-D_\nu A^a_\mu,\\
A^a_{b\mu\nu}&=&D_\mu A^a_{b\nu}-D_\nu A^a_{b\mu},\\
D_\mu A_\nu^a&=&\partial_\mu A_\nu^a-iV_\mu A_\nu^a-iA_\nu^a V_\mu^T,\\
D_\mu S^a&=&\partial_\mu S^a-iV_\mu S^a-iS^aV_\mu^T,\\
D_\mu A_{b\nu}^a&=&\partial_\mu A_{b\nu}^a-iA_{b\nu}^a\alpha_{\|\mu}^T,\\
D_\mu S_b^a&=&\partial_\mu S_b^a-iS_b^a\alpha_{\|\mu}^T.
\end{eqnarray}
In Eqs. (\ref{eqLagrangian2}) and (\ref{eqLagrangian3}), the Einstein summation convention is used, i.e., the repeated superscripts ``$a$'' mean the summation over them.
For $\mathcal{L}_1$, the constants $\beta=0.9$ and $g=0.59$. For $\mathcal{L}_2$ and $\mathcal{L}_3$, there are two set of coupling constants $e_i\ (i=1,2,...,9)$ and $h_j\ (j=1,2,...,5)$ whose values are still unknown. In this work, we naively use the $^3P_0$ model to determine them. Their values are listed in Table I. Note that we can not fix the sign of $e_3$, $e_5$ and $e_6$ because the relative phase between the amplitudes obtained from the Lagrangian and $^3P_0$ model can not be determined. For $e_7$, $e_8$ and $e_9$, we use the phase in Ref. \cite{Cao:2022rjp} which explains the $Z_{cs}$ well and get the values of them. In Eqs. \eqref{eq21} and \eqref{eq22}, we choose $\sigma=0$ according to the unitary gauge \cite{Harada:2003jx}.
\begin{table}
\caption{The values of the low energy constants in the Lagrangians containing diquarks.}\label{tab1}
\begin{tabular}{ccccc}\toprule[1pt]
$e_1$ (GeV$^{-1}$)&$e_2$ (GeV$^{-1}$)&$e_3$ (GeV$^{-2}$)&$e_4$ (GeV$^{-1}$)&$e_5$ (GeV$^{-2}$)\\\midrule[0.5pt]
-6.353&1.657&$\pm$0.555&4.885&$\pm$0.566\\\midrule[0.5pt]
$e_6$ (GeV$^{-2}$)&$e_7$ (GeV$^{-1}$)&$e_8$ (GeV$^{-1}$)&$e_9$ (GeV$^{-1}$)&\\\midrule[0.5pt]
$\pm$1.005&-0.909&-13.348& 11.530\\\midrule[0.5pt]
$h_1$ (GeV$^{-1}$)&$h_2$ (GeV$^{-1}$)&$h_3$ (GeV$^{-1}$)&$h_4$ (GeV$^{-1}$)&$h_5$ (GeV$^{-1}$)\\\midrule[0.5pt]
0.084&-0.130&0.266&1.457&0.011
\\\bottomrule[1pt]
\end{tabular}
\end{table}

\subsection{Effective Potentials With The Exponential Form Factor}\label{subsection: effective potentials}
Making use of the Breit approximation, we obtain the effective potentials in the momentum space
\begin{equation}
    \mathcal{V}^{H_1H_2\to H_3 H_4}(\bm{q})=\frac{\mathcal{M}^{H_1H_2\to H_3 H_4}(\bm{q})}{\sqrt{\prod_{i}2m_{i}\prod_{f}2m_f}}
\end{equation}
where $m_i\ (i=1,2,3,4)$ denotes the mass of the particle labeled by $i$. By performing the Fourier transformation, we get the effective potential in the coordinate space
\begin{equation}
    \small
    \mathcal{V}^{H_1H_2\to H_3 H_4}(\bm{r})=\int\frac{d^{3}\bm{q}}{(2\pi)^3}e^{i\bm{q\cdot r}}\mathcal{V}^{H_1H_2\to H_3 H_4}(\bm{q})F^{2}(q^2).
\end{equation}
Here, $F(\Vec{q}^2)$ is the form factor which suppresses the contribution of high momenta, i.e., small distance. And the presence of such a form factor is dictated by the extended (quark) structure of the hadrons. In this work, we adopt the exponentially parameterized form factor
\begin{equation}
    F(q^2)=e^{q^2/\Lambda^2}=e^{(q_{0}^{2}-\vec{q}^2)/\Lambda^2}
\end{equation}
with $\Lambda$ the cut-off. 

Another option of the form factor is monopole expression, i.e.,
\begin{eqnarray}
F_M(q^2)=\frac{\Lambda^2-m_{E}^2}{\Lambda^2-q^2}
\end{eqnarray}
with $m_{E}$ the mass of the exchanged particle. If the exchanged meson's mass is large, for instance in the case of $\phi$- or diquark-exchange, $F_M(q^2)$ is highly suppressed by the numerator $\Lambda^2-m_{E}^2$, which would lead to unreasonable results. Consequently, we choose the exponentially parameterized form factor in this work.

In Eqs. (\ref{eq33}-\ref{eq51}), we list the specific expressions of the non-zero effective subpotentials which are isospin-independent:
{
\footnotesize
\begin{eqnarray}
V^{\bar{B}^{*}B\to \bar{B}^{*}B/\bar{B}B^{*}\to \bar{B}B^{*}}_{v}&=&C_{v}\frac{(\beta g_{V})^{2}m_{B^{*}}m_{B}}{4M_{P}M_{P^{*}}}(\bm{\epsilon_{1/2}}\cdot \bm{\epsilon^{\dagger}_{3/4}})\nonumber\\
&&\times Y(\Lambda,m_{v},r), \label{eq33}\\
V^{\bar{B}B^{*}\to \bar{B}^{*}B}_{p}&=&C_{p}\left(\frac{g}{2F_{\pi}}\right)^{2}\bigg[(\bm{\epsilon_{2}}\cdot \bm{\epsilon^{\dagger}_{3}})Z(\Lambda,\Tilde{m}_{p},r)\nonumber\\
&&+S(\hat{\bm{r}},\bm{\epsilon_{2}},\bm{\epsilon^{\dagger}_{3}})T(\Lambda,\Tilde{m}_{p},r)\bigg],\\
V^{\bar{B}^{*}B/\bar{B}B^{*}\to \bar{B}^{*}B^*}_{p}&=&\delta_{B\bar{B}^{*}/B^{*}\bar{B}}C_{p}\frac{g^2m_{B^*}}{4M_{P^*}F^{2}_{\pi}}\nonumber\\
&&\times \bigg[\bm{\epsilon^{\dagger}_{4/3}}\cdot(\bm{\epsilon_{1/2}}\times\bm{\epsilon^{\dagger}_{3/4})}Z(\Lambda,\Tilde{m}_{p},r)\nonumber\\
&&+S(\hat{\bm{r}},\bm{\epsilon^{\dagger}_{4/3}},\bm{\epsilon_{1/2}\times\epsilon^{\dagger}_{3/4}})\nonumber\\
&&\times T(\Lambda,\Tilde{m}_{p},r)\bigg], \\
V^{\bar{B}^{*}B\to S\bar{A}/\bar{B}B^{*}\to A\bar{S}}_{S_{ud}}&=&-\frac{\sqrt{3}}{12}e_{1}e_{4}\bigg[(\bm{\epsilon_{1/2}}\cdot \bm{\epsilon^{\dagger}_{4/3}})Z(\Lambda,\Tilde{m}_{S_{ud}},r)\nonumber\\
&&+S(\hat{\bm{r}},\bm{\epsilon_{1/2}}, \bm{\epsilon^{\dagger}_{4/3}})T(\Lambda,\Tilde{m}_{S_{ud}},r)\bigg],\\
V^{\bar{B}^{*}B\to S\bar{A}/\bar{B}B^{*}\to A\bar{S}}_{A_{ud}}&=&\sqrt{3}e_{3}e_{6}m_{S_{bq}}m_{A_{bq}}\bigg[\frac{2}{3}(\bm{\epsilon_{1/2}}\cdot \bm{\epsilon^{\dagger}_{4/3}})\nonumber\\
&&\times Z(\Lambda,\Tilde{m}_{A_{ud}},r)-\frac{1}{3}S(\hat{\bm{r}},\bm{\epsilon_{1/2}}, \bm{\epsilon^{\dagger}_{4/3}})\nonumber\\
&&\times T(\Lambda,\Tilde{m}_{A_{ud}},r)\bigg],\\
%
%
V^{\bar{B}^{*}B\to A\bar{S}/\bar{B}B^{*}\to S\bar{A}}_{A_{ud}}&=&-\frac{\sqrt{3}}{8}e_{2}(e_{8}m_{A_{bq}}+e_{9}m_{B^{*}})m_{S_{bq}}\nonumber\\
&&\times \left(\frac{\Tilde{m}_{A_{ud}}}{m_{A_{ud}}}\right)^{2}(\bm{\epsilon_{1/2}}\cdot \bm{\epsilon^{\dagger}_{3/4}})\nonumber\\
&&\times Y(\Lambda,\Tilde{m}_{A_{ud}},r),\\
V^{\bar{B}^{*}B/\bar{B}B^{*}\to A\bar{A}}_{S_{ud}}&=&\delta_{B\bar{B}^{*}/B^{*}\bar{B}}\frac{1}{2\sqrt{3}}e_{1}e_{5}m_{A_{bq}}\nonumber\\
&&\times \bigg[\bm{\epsilon^{\dagger}_{4/3}}\cdot(\bm{\epsilon_{1/2}}\times\bm{\epsilon^{\dagger}_{3/4}})Z(\Lambda,\Tilde{m}_{S_{ud}},r)\nonumber\\
&&+S(\hat{\bm{r}},\bm{\epsilon^{\dagger}_{4/3}},\bm{\epsilon_{1/2}}\times\bm{\epsilon^{\dagger}_{3/4}})\nonumber\\
&&\times T(\Lambda,\Tilde{m}_{S_{ud}},r)\bigg],\\ 
V^{\bar{B}^{*}B/\bar{B}B^{*}\to A\bar{A}}_{A_{ud}}&=&\delta_{B\bar{B}^{*}/B^{*}\bar{B}}\sqrt{\frac{1}{3}}e_{3}e_{7}m_{A_{bq}}\nonumber\\
&&\times \bigg[\bm{\epsilon_{1/2}}\cdot(\bm{\epsilon^{\dagger}_{3/4}}\times\bm{\epsilon^{\dagger}_{4/3}})Z(\Lambda,\Tilde{m}_{A_{ud}},r)\nonumber\\
&&-\frac{1}{2}S(\hat{\bm{r}},\bm{\epsilon^{\dagger}_{3/4}},\bm{\epsilon_{1/2}}\times\bm{\epsilon^{\dagger}_{4/3}})\nonumber\\
&&\times T(\Lambda,\Tilde{m}_{A_{ud}},r)\nonumber\\
&&+\frac{1}{2}S(\hat{\bm{r}},\bm{\epsilon_{1/2}},\bm{\epsilon^{\dagger}_{3/4}}\times\bm{\epsilon^{\dagger}_{4/3}})\nonumber
\end{eqnarray}
\begin{eqnarray}
&&\times T(\Lambda,\Tilde{m}_{A_{ud}},r)\bigg],\\ 
V^{\bar{B}^{*}B^*\to \bar{B}^{*}B^*}_{v}&=&C_v\left(\frac{\beta g_V m_{B^{*}}}{2M_{P^{*}}}\right)^{2}(\bm{\epsilon_{1}}\cdot\bm{\epsilon^{\dagger}_{3}})(\bm{\epsilon_{2}}\cdot\bm{\epsilon^{\dagger}_{4}})\nonumber\\
&&\times Y(\Lambda,m_{v},r), \label{eq40}\\
V^{\bar{B}^{*}B^*\to \bar{B}^{*}B^*}_{p}&=&C_{p}\left(\frac{g m_{B^{*}}}{2m_{P^{*}F_{\pi}}}\right)^{2}\bigg[(\bm{\epsilon_{1}}\times\bm{\epsilon^{\dagger}_{3}})\cdot(\bm{\epsilon_{2}}\times\bm{\epsilon^{\dagger}_{4}})\nonumber\\
&&\times Z(\Lambda,m_{p},r)+S(\hat{\bm{r}},\bm{\epsilon_{1}}\times\bm{\epsilon^{\dagger}_{3}},\bm{\epsilon_{2}}\times\bm{\epsilon^{\dagger}_{4}})\nonumber\\
&&\times T(\Lambda,m_{p},r)\bigg],\\
V^{\bar{B}^{*}B^*\to S\bar{A}/\bar{S}A}_{S_{ud}}&=&-\frac{\delta_{S\bar{A}/A\bar{S}}}{2\sqrt{3}}e_{4}e_{5}m_{A_{bq}}\bigg[\bm{\epsilon_{1/2}}\cdot(\bm{\epsilon_{2/1}}\times\bm{\epsilon^{\dagger}_{4/3}})\nonumber\\
&&\times Z(\Lambda,\Tilde{m}_{S_{ud}},r)+S(\hat{\bm{r}},\bm{\epsilon_{1/2}},\bm{\epsilon_{2/1}}\times\bm{\epsilon^{\dagger}_{4/3}})\nonumber\\
&&\times T(\Lambda,\Tilde{m}_{S_{ud}},r),\\
V^{\bar{B}^{*}B^*\to S\bar{A}/A\bar{S}}_{A_{ud}}&=&-\delta_{S\bar{A}/A\bar{S}}\frac{1}{2\sqrt{3}}e_{6}e_{7}m_{S_{bq}}\nonumber\\
&&\times\bigg[\bm{\epsilon_{2/1}}\cdot(\bm{\epsilon_{1/2}}\times\bm{\epsilon^{\dagger}_{4/3}}) Z(\Lambda,\Tilde{m}_{A_{ud}},r)\nonumber\\
&&+\frac{1}{2}S(\hat{\bm{r}},\bm{\epsilon_{2/1}},\bm{\epsilon_{1/2}}\times\bm{\epsilon^{\dagger}_{4/3}})T(\Lambda,\Tilde{m}_{A_{ud}},r)\nonumber\\
&& -\frac{1}{2}S(\hat{\bm{r}},\bm{\epsilon^{\dagger}_{4/3}},\bm{\epsilon_{1/2}}\times\bm{\epsilon_{2/1}})T(\Lambda,\Tilde{m}_{A_{ud}},r)\bigg],\nonumber\\ \\
V^{\bar{B}^{*}B^*\to A\bar{A}}_{S_{ud}}&=& \frac{e^2_5m^{2}_{A_{bq}}}{\sqrt{3}}\bigg[(\bm{\epsilon_1}\times\bm{\epsilon^\dagger_3})\cdot(\bm{\epsilon_2}\times\bm{\epsilon^\dagger_4})Z(\Lambda,\Tilde{m}_{S_{ud}},r)\nonumber\\
&&+S(\bm{r},\bm{\epsilon_1}\times\bm{\epsilon^\dagger_3},\bm{\epsilon_2}\times\bm{\epsilon^\dagger_4})T(\Lambda,\Tilde{m}_{S_{ud}},r)\bigg],\\
V^{\bar{B}^{*}B^*\to A\bar{A}}_{A_{ud}}&=&-\frac{\sqrt{3}e^2_7}{24}\bigg[2(\bm{\epsilon_2}\cdot\bm{\epsilon^\dagger_3})(\bm{\epsilon_1}\cdot\bm{\epsilon^\dagger_4})Z(\Lambda,\Tilde{m}_{A_{ud}},r)\nonumber\\
&&-2(\bm{\epsilon_1\cdot\epsilon_2})(\bm{\epsilon^\dagger_3}\cdot\bm{\epsilon^\dagger_4})Z(\Lambda,\Tilde{m}_{A_{ud}},r)\nonumber\\
&&+(\bm{\epsilon_2}\cdot\bm{\epsilon^\dagger_3})S(\bm{r},\bm{\epsilon_1,\epsilon^\dagger_4})T(\Lambda,\Tilde{m}_{A_{ud}},r)\nonumber\\
&&-(\bm{\epsilon^\dagger_3}\cdot\bm{\epsilon^\dagger_4})S(\hat{\bm{r}},\bm{\epsilon_1},\bm{\epsilon_2})T(\Lambda,\Tilde{m}_{A_{ud}},r)\nonumber\\
&&+\bm{(\epsilon_1\cdot\epsilon^\dagger_4)}S(\hat{\bm{r}},\bm{\epsilon_2},\bm{\epsilon^\dagger_3})T(\Lambda,\Tilde{m}_{A_{ud}},r)\nonumber\\
&&-(\bm{\epsilon_1}\cdot\bm{\epsilon_2})S(\hat{\bm{r}},\bm{\epsilon^\dagger_3},\bm{\epsilon^\dagger_4})T(\Lambda,\Tilde{m}_{A_{ud}},r)\bigg]\nonumber\\
&&-\frac{\sqrt{3}}{8}(e_{8}m_{A_{bq}}+e_{9}m_{B^{*}})^{2}\left(\frac{\Tilde{m}_{A_{ud}}}{m_{A_{ud}}}\right)^{2}\nonumber\\
&&\times (\bm{\epsilon_1}\cdot\bm{\epsilon^\dagger_3})(\bm{\epsilon_2}\cdot\bm{\epsilon^\dagger_4})Y(\Lambda,\Tilde{m}_{A_{ud}},r),\\
V^{S\bar{A}\to S\bar{A}/A\bar{S}\to A\bar{S}}_{v}&=&C_v\left(\frac{g_V}{2}\right)^{2}h_{1}h_{4}m_{A_{bq}}m_{S_{bq}}(\bm{\epsilon_{2/1}}\cdot \bm{\epsilon^{\dagger}_{4/3}})\nonumber\\
&&\times Y(\Lambda,m_{v},r),\\
V^{S\bar{A}\to A\bar{S}/A\bar{S}\to S\bar{A}}_{p}&=&C_{p}\left(\frac{h_3}{4F_{\pi}}\right)^{2}\bigg[(\bm{\epsilon_{1/2}}\cdot \bm{\epsilon^{\dagger}_{4/3}})Z(\Lambda,\Tilde{m}_{p},r)\nonumber\\
&&+S(\hat{\bm{r}},\bm{\epsilon_{1/2}},\bm{\epsilon^{\dagger}_{4/3}})T(\Lambda,\Tilde{m}_{p},r)\bigg],\\
V^{S\bar{A}/A\bar{S}\to A\bar{A}}_{p}&=&\delta_{S\bar{A}/A\bar{S}}C_{p}\frac{h_3 h_5}{4F^{2}_{\pi}}m_{A_{bq}}\nonumber\\
&&\times\bigg[\bm{\epsilon^{\dagger}_{3/4}}\cdot(\bm{\epsilon_{2/1}}\times\bm{\epsilon^{\dagger}_{4/3}}) Z(\Lambda,\Tilde{m}_{p},r)\nonumber\\
&&+S(\hat{\bm{r}},\bm{\epsilon^{\dagger}_{3/4}},\bm{\epsilon_{2/1}}\times\bm{\epsilon^{\dagger}_{4/3}})T(\Lambda,\Tilde{m}_{p},r)\bigg],\\
V^{A\bar{A}\to A\bar{A}}_{v}&=&C_v\frac{g^2_V h^2_4m^2_{A_{bq}}}{4}(\bm{\epsilon_1}\cdot\bm{\epsilon^{\dagger}_{3}})(\bm{\epsilon_2}\cdot\bm{\epsilon^{\dagger}_{4}})Y(\Lambda,m_{v},r),\nonumber\\
\end{eqnarray}
\begin{eqnarray} 
V^{A\bar{A}\to A\bar{A}}_{p}&=&C_{p}\left(\frac{h_{5}m_{A_{bq}}}{F_\pi}\right)^2 \bigg[(\bm{\epsilon_1}\times\bm{\epsilon^{\dagger}_{3}})\cdot(\bm{\epsilon_2\times\epsilon^{\dagger}_{4}})\nonumber\\ 
&&\times Z(\Lambda,m_{p},r)+S(\bm{r},\bm{\epsilon_1}\times\bm{\epsilon^{\dagger}_{3}},\bm{\epsilon_2}\times\bm{\epsilon^{\dagger}_{4}})\nonumber\\
&&\times T(\Lambda,m_{p},r)\bigg].\label{eq51}
\end{eqnarray}
}
In the above equations, the subscript $p$ denotes the light pseudoscalar mesons $\pi$, $\eta_8$ and $\eta_0$, and $v$ denotes the light vectors $\rho$ and $\omega$. The coefficients $C_{\pi^0}=\frac{1}{3}, C_{\eta_8}=-\frac{1}{9}, C_{\eta_0}=-\frac{2}{9}$, $C_{\rho}=1$, $C_\omega=-1$, $\delta_{B\Bar{B}^*}=1$, $\delta_{B^*\Bar{B}}=-1$, $\delta_{S\Bar{A}}=1$, $\delta_{A\Bar{S}}=-1$.
And $S(\hat{\bm r},\bm{a},\bm{b})=3(\hat{\bm r}\cdot\bm{a})(\hat{\bm r}\cdot\bm{b})-\bm{a\cdot b}$, $q_0=\frac{m^2_2-m^2_1+m^2_3-m^2_4}{2(m_3+m_4)}$, $\Tilde{m}^2_{E}=m^2_{E}-q_{0}^{2}$ with $E$ the exchanged particle. The functions $Y(\Lambda,m,r)$, $Z(\Lambda,m,r)$ and $T(\Lambda,m,r)$ are defined as 
\begin{eqnarray}
Y(\Lambda,m,r)&=&\int \frac{d^3q}{(2\pi)^3}e^{i\vec{q}\cdot \vec{r}}\frac{1}{\vec{q}^2+m^2-i\epsilon}e^{2(q_0^2-\vec{q}^2)/\Lambda^2},\nonumber\\\label{eq52}\\
Z(\Lambda,m,r)&=&\frac{1}{r^2}\frac{\partial}{\partial r}r^{2}\frac{\partial}{\partial r}Y(\Lambda,m,r),\\
T(\Lambda,m,r)&=&r\frac{\partial}{\partial r}\frac{1}{r}\frac{\partial}{\partial r}Y(\Lambda,m,r).
\end{eqnarray}
In the Appendix, we will show the calculation method of the integral $Y(\Lambda,m,r)$.

Taking into account S- and D-wave functions, the products of the polarization vectors in the subpotentials are presented below
\begin{eqnarray}
\left.
\begin{array}{cc}
\bm{\epsilon_{1}\cdot\epsilon^{\dagger}_{3/4}}\\
\bm{\epsilon_{2}\cdot\epsilon^{\dagger}_{3/4}}\\
(\bm{\epsilon_{1}\cdot\epsilon^{\dagger}_{3}})(\bm{\epsilon_{2}\cdot\epsilon^{\dagger}_{4}})\\
-(\bm{\epsilon_{1}\cdot\epsilon^{\dagger}_{4}})(\bm{\epsilon_{2}\cdot\epsilon^{\dagger}_{3}})\\(\bm{\epsilon_{1}\times\epsilon^{\dagger}_{3}})\cdot(\bm{\epsilon_{2}\times\epsilon^{\dagger}_{4}})
\end{array}\right\}
&\longrightarrow& \left(
\begin{array}{cc}
    1 & 0 \\
    0 & 1
\end{array}\right),\nonumber\\
\left.
\begin{array}{cc}
S(\hat{\bm{r}},\bm{\epsilon_1},\bm{\epsilon_4}^\dag)\\
S(\hat{\bm{r}},\bm{\epsilon_2},\bm{\epsilon_3}^\dag)\\
S(\hat{\bm{r}},\bm{\epsilon_{1}\times\epsilon^{\dagger}_{3}},\bm{\epsilon_{2}\times\epsilon^{\dagger}_{4}})\\
2(\bm{\epsilon_{2}\cdot\epsilon^{\dagger}_{3}})S(\hat{\bm{r}},\bm{\epsilon_{1},\epsilon^{\dagger}_{4}})\\
2(\bm{\epsilon_{1}\cdot\epsilon^{\dagger}_{4}})S(\hat{\bm{r}},\bm{\epsilon_{2},\epsilon^{\dagger}_{3}})
\end{array}\right\}
&\longrightarrow& {\left(
\begin{array}{cc}
    0 & -\sqrt{2} \\
    -\sqrt{2} & 1
\end{array}\right)},\nonumber\\
\left.
\begin{array}{cc}
\bm{\epsilon_{1/2}}\cdot(\bm{\epsilon^{\dagger}_{4}\times\epsilon^{\dagger}_{3}})\\
\bm{\epsilon_{1}}\cdot(\bm{\epsilon_{2}\times\epsilon^{\dagger}_{3/4}})          
\end{array}\right\}
&\longrightarrow& \left(
\begin{array}{cc}
    i\sqrt{2} & 0 \\
    0 & i\sqrt{2}
\end{array}\right),\nonumber\\
\left.
\begin{array}{cc}
2S(\hat{\bm{r}},\bm{\epsilon^{\dagger}_{4}},\bm{\epsilon^{\dagger}_{3}\times\epsilon_{1/2}})\\
2S(\hat{\bm{r}},\bm{\epsilon^{\dagger}_{3}},\bm{\epsilon_{1/2}\times\epsilon^{\dagger}_{4}})\\
2S(\hat{\bm{r}},\bm{\epsilon_{1}},\bm{\epsilon_2\times\epsilon^{\dagger}_{3/4}})\\
2S(\hat{\bm{r}},\bm{\epsilon_{2}},\bm{\epsilon^{\dagger}_{3/4}\times\epsilon_{1}}) \\
S(\hat{\bm{r}},\bm{\epsilon_{1/2}},\bm{\epsilon^{\dagger}_{3}\times\epsilon^{\dagger}_{4}})\\
S(\hat{\bm{r}},\bm{\epsilon^{\dagger}_{3/4}},\bm{\epsilon_{2}\times\epsilon_{1}})
\end{array}\right\}
&\longrightarrow& \left(
\begin{array}{cc}
    0 & 2i \\
    2i & -i\sqrt{2}
\end{array}\right),\nonumber\\
\left.
\begin{array}{cc} 
\bm{\epsilon_{1}\cdot\epsilon_{2}},\bm{\epsilon^{\dagger}_{3}\cdot\epsilon^{\dagger}_{4}}\\
(\bm{\epsilon^{\dagger}_{3}\cdot\epsilon^{\dagger}_{4}})S(\hat{\bm{r}},\bm{\epsilon_{1},\epsilon_{2}})\\
(\bm{\epsilon_{1}\cdot\epsilon_{2}})S(\hat{\bm{r}},\bm{\epsilon^{\dagger}_{3},\epsilon^{\dagger}_{4}})
\end{array}\right\}
&\longrightarrow & \left(
\begin{array}{cc}
    0 & 0 \\
    0 & 0
\end{array}\right).\nonumber
\end{eqnarray}
Hereafter, we label the channels $B\Bar{B}^*/B^*\Bar{B}$, $B^*\Bar{B}^*$, $S\Bar{A}/A\Bar{S}$ and $A\Bar{A}$ by $CH_1$, $CH_2$, $CH_3$ and $CH_4$, respectively. The elements of the total potential matrix $\hat{V}$ is 
{\small
\begin{eqnarray}
V^{CH_1\to CH_1}&=&\frac{1}{2}\left(V^{\bar{B}B^*\to \bar{B}B^*}_{\rho,\omega}+2V^{\bar{B}B^*\to \bar{B}^*B}_{\pi,\eta_8,\eta_0}\right.\nonumber\\
&&\left.+V^{\bar{B}^*B\to \bar{B}^*B}_{\rho,\omega}\right),\\
V^{CH_1\to CH_2}&=&\frac{1}{\sqrt{2}}\left(V^{\bar{B}^*B\to \bar{B}^*B^*}_{\pi,\eta_8,\eta_0}+V^{\bar{B}B^*\to \bar{B}^*B^*}_{\pi,\eta_8,\eta_0}\right),\\
V^{CH_1\to CH_3}&=&\frac{1}{2}\left(V^{\bar{B}^*B\to A\Bar{S}}_{A_{ud}}+V^{\bar{B}B^*\to A\Bar{S}}_{S_{ud},A_{ud}}\right.\nonumber\\
&&\left.+V^{\bar{B}^*B\to S\Bar{A}}_{S_{ud},A_{ud}}+V^{\bar{B}^*B\to A\Bar{S}}_{A_{ud}}\right),\\
V^{CH_1\to CH_4}&=&\frac{1}{\sqrt{2}}\left(V^{\bar{B}^*B\to A\Bar{A}}_{S_{ud},A_{ud}}+V^{\bar{B}B^*\to A\Bar{A}}_{S_{ud},A_{ud}}\right),\\
V^{CH_2\to CH_2}&=&V^{\bar{B}^*B^*\to \bar{B}^*B^*}_{\rho,\omega}+V^{\bar{B}^*B^*\to \bar{B}^*B^*}_{\pi,\eta_8,\eta_0},\label{eq57}\\
V^{CH_2\to CH_3}&=&\frac{1}{\sqrt{2}}\left(V^{\bar{B}^*B^*\to A\Bar{S}}_{S_{ud},A_{ud}}+V^{\bar{B}^*B^*\to S\Bar{A}}_{S_{ud},A_{ud}}\right),\\
V^{CH_2\to CH_4}&=&V^{\bar{B}^*B^*\to A\Bar{A}}_{S_{ud},A_{ud}},\\
V^{CH_3\to CH_3}&=&\frac{1}{2}\left(V^{S\Bar{A}\to S\Bar{A}}_{\rho,\omega}+2V^{S\Bar{A}\to A\Bar{S}}_{\pi,\eta_8,\eta_0}\right.\nonumber\\
&&\left.+V^{A\Bar{S}\to A\Bar{S}}_{\rho,\omega}\right),\\
V^{CH_3\to CH_4}&=&\frac{1}{\sqrt{2}}\left(V^{S\bar{A}\to A\Bar{A}}_{\pi,\eta_8,\eta_0}+V^{A\bar{S}\to A\Bar{A}}_{\pi,\eta_8,\eta_0}\right),\\
V^{CH_4\to CH_4}&=&V^{A\bar{A}\to A\Bar{A}}_{\rho,\omega}+V^{A\bar{A}\to A\Bar{A}}_{\pi,\eta_8,\eta_0}.
\end{eqnarray}
}
The subscripts mean the following summation
\begin{eqnarray}
V^{CH_i\to CH_j}_{E_1,E_2,\cdots}=\sum_{a=E_1,E_2,\cdots}V^{CH_i\to CH_j}_{a}.    
\end{eqnarray}

\section{Results and Discussion}\label{results}

With the preparation above, adopting the Gaussian expansion method (GEM) \cite{Hiyama:2003cu}, we solve the coupled channel Schr\"{o}dinger equation to find the bound state solutions, 
\begin{align}
    \left(\hat{K}+\hat{M}+\hat{V}\right)\Psi=E\Psi.\label{eq66}
\end{align}
Here, $\hat{K}=\text{diag}(-\frac{\Delta}{2\mu_1},-\frac{\Delta}{2\mu_2},\cdots)$, $\hat{M}=\text{diag}(0,M_2-M_1,M_3-M_1,\cdots)$. In the central force field problem, the system does not depend on the azimuth and polar angle, so the operator $\Delta=\frac{1}{r^2}\frac{\partial}{\partial r}\left(r^2\frac{\partial}{\partial r}\right)$.
The coupled channel Schr\"{o}dinger equation Eq. (\ref{eq66}) is symmetric under the following transformation
\begin{eqnarray}
&&U\left(\hat{K}+\hat{M}+\hat{V}\right)\Psi=UE\Psi,\\
&\Rightarrow&U\left(\hat{K}+\hat{M}+\hat{V}\right)U^{-1}U\Psi=EU\Psi,\\
&\Rightarrow&\left(\hat{K}+\hat{M}+U\hat{V}U^{-1}\right)\Tilde{\Psi}=E\Tilde{\Psi},
\end{eqnarray}
where $U=\text{diag}(e^{i\theta_1},e^{i\theta_2},\cdots)$, $\Tilde{\Psi}=U\Psi$. Here the word ``symmetric'' means the energy $E$ does not change under the transformation of the Schr\"{o}dinger equation. All the $U$ form a reducible Lie group $U(1)\otimes U(1)\otimes \cdots\otimes U(1)$. 

In this work, some of the off-diagonal elements of the potential matrix are imaginary, which is not convenient for solving the Schr\"{o}dinger equation. If we perform the transformation mentioned above, the problem is solved which means that the potential matrix becomes real. The corresponding parameters are chosen as $\theta_1=\frac{\pi}{2}$, $\theta_2=0$, $\theta_3=\frac{\pi}{2}$, $\theta_4=0$, $\cdots$.

\begin{figure}[htbp]
    \centering
    \includegraphics[width=1.0\linewidth]{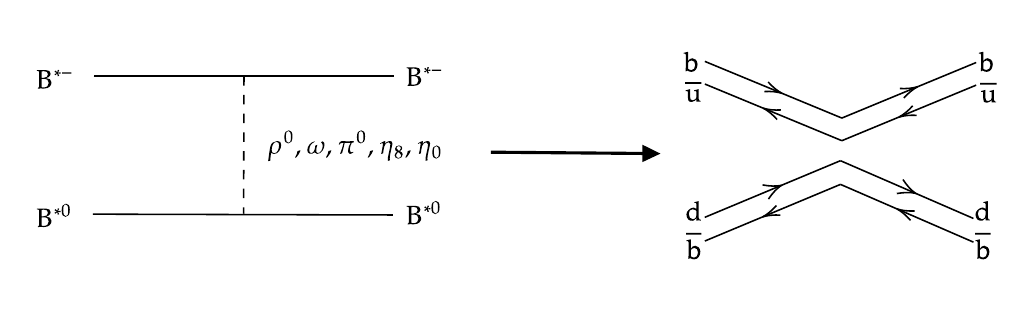}
    \caption{The diagram of $B^{*-}B^{*0}\to B^{*-}B^{*0}$ on quark and hadron levels.}
    \label{fig: scattering processes}
\end{figure} 
It is also interesting to discuss the OZI suppressed processes on both the quark and hadron levels, which was already pointed in Ref. \cite{Dias:2014pva}. We take the $B^{*-}\Bar{B}^{*0}\to B^{*-}\Bar{B}^{*0}$ process as an example. On the quark level, we can see from Fig. \ref{fig: scattering processes} that the diagram is non-connected, which means it is OZI suppressed. On the hadron level, the exchanged particle could be the pseudoscalar mesons $\pi, \eta_8, \eta_0$ and the vector mesons $\rho, \omega$. Since the masses of $\rho$ and $\omega$ are almost close to each other, their contributions to the total potential cancel out, which can be seen from Eqs. (\ref{eq40}) and (\ref{eq57}). Under the $[U(3)_L\otimes U(3)_R]_{global}\otimes [U(3)]_{local}$ symmetry, the pseudoscalar mesons appearing in $\Phi$ (see Eq. (\ref{eq7})) have the same mass. In this case, The contributions of $\pi$, $\eta_8$ and $\eta_0$ to the total potential also cancel out. This conclusion coincides with the OZI rule. However, the real masses of $\pi$, $\eta_8$ and $\eta_0$ are different, so the total potential is non-zero which depicts the phenomenon of the violation of the OZI rule.
\begin{table}
\caption{The diquark masses we use in effective potentials from Ref. \cite{Ferretti:2019zyh}. $q$ indicates $u$ or $d$ quark.}\label{tab3}
\begin{tabular}{ccccc}\toprule[1pt]
$m_{S_{bq}}$ (GeV)&$m_{A_{bq}}$ (GeV)&$m_{S_{qq}}$ (GeV)&$m_{A_{qq}}$ (GeV)\\\midrule[0.5pt]
5.451&5.465&0.691&0.840
\\\bottomrule[1pt]
\end{tabular}
\end{table}

\begin{figure*}[htb]
    \centering
    \includegraphics[width=0.3\linewidth]{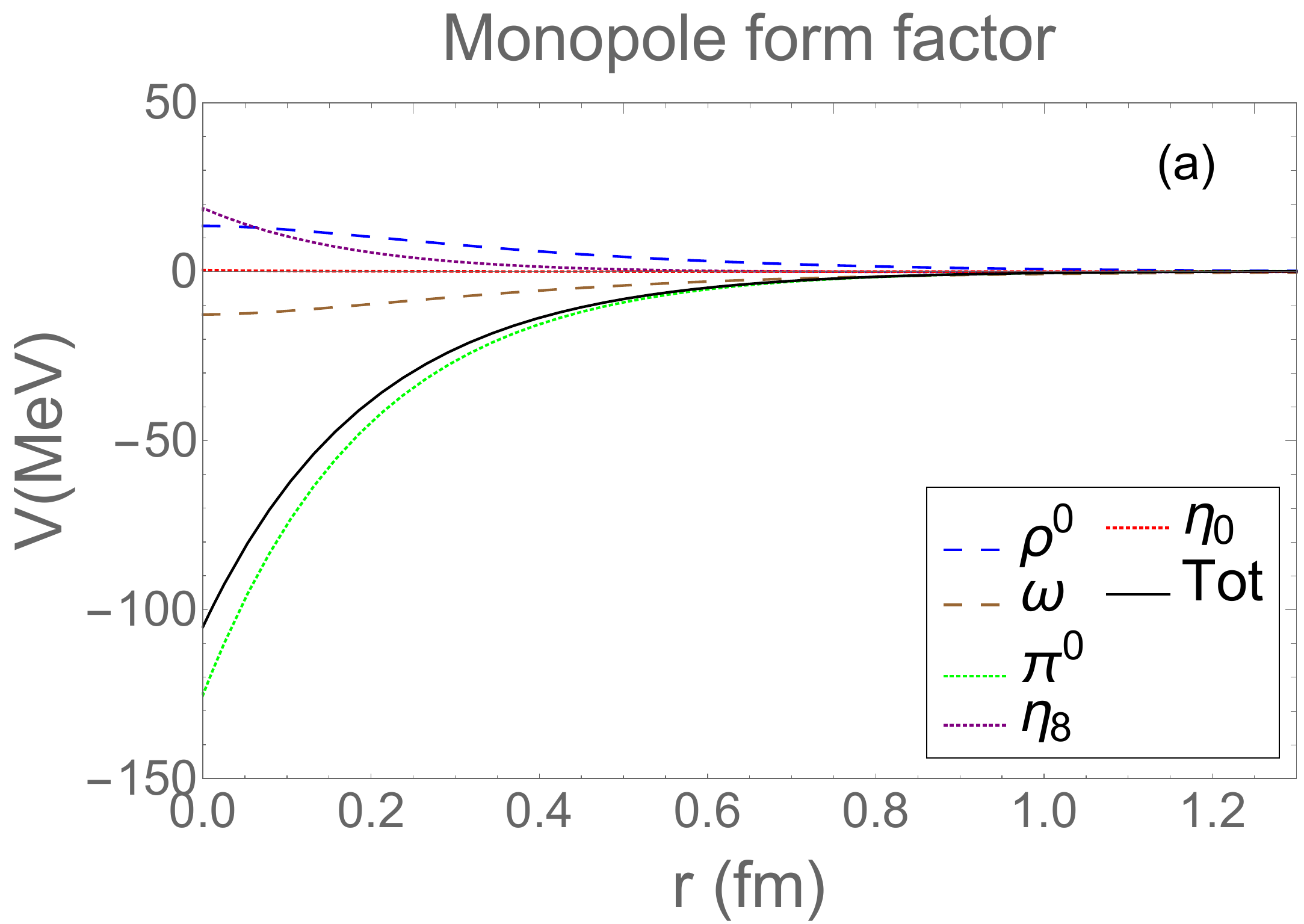}\includegraphics[width=0.3\linewidth]{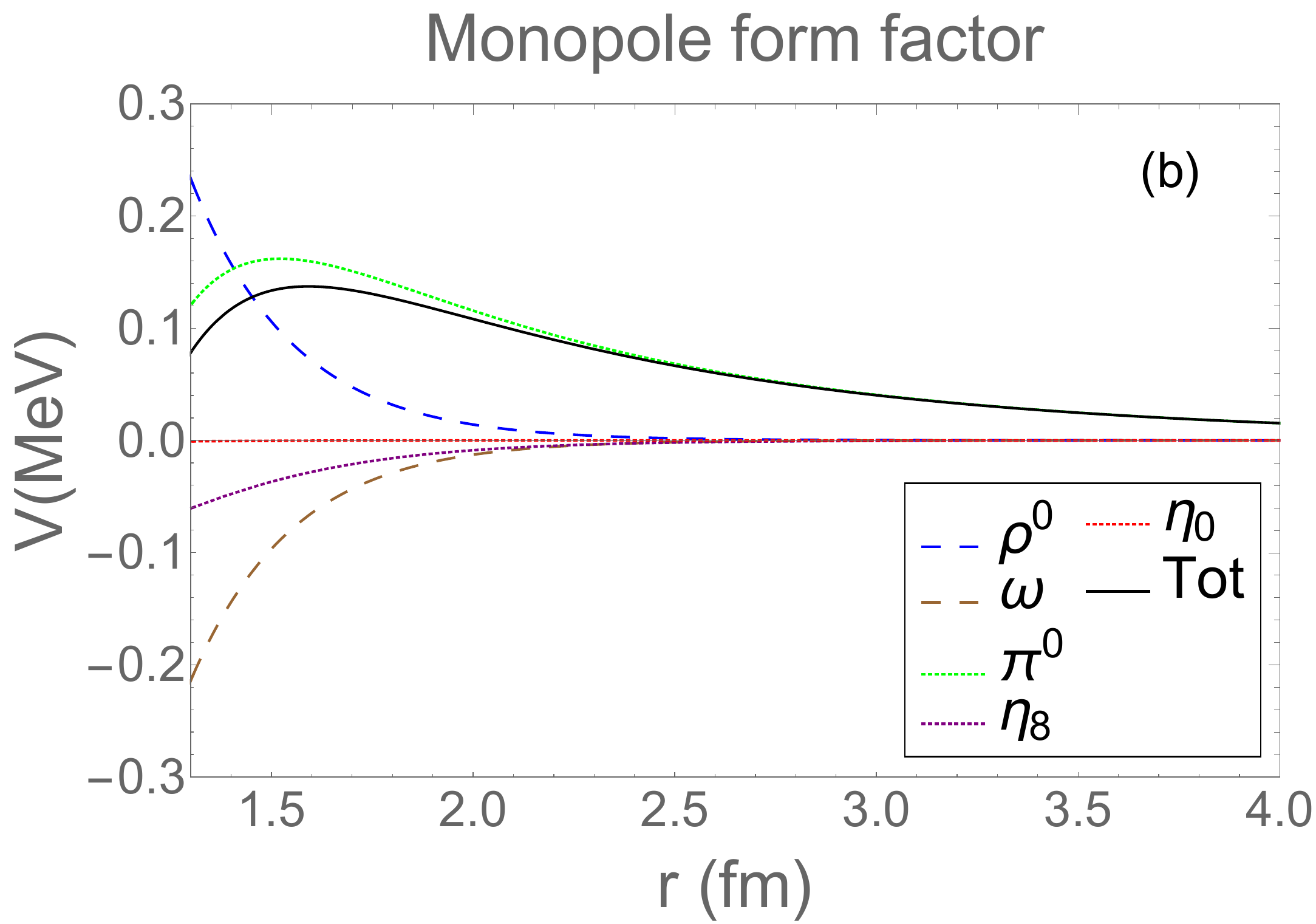}\\
    \includegraphics[width=0.3\linewidth]{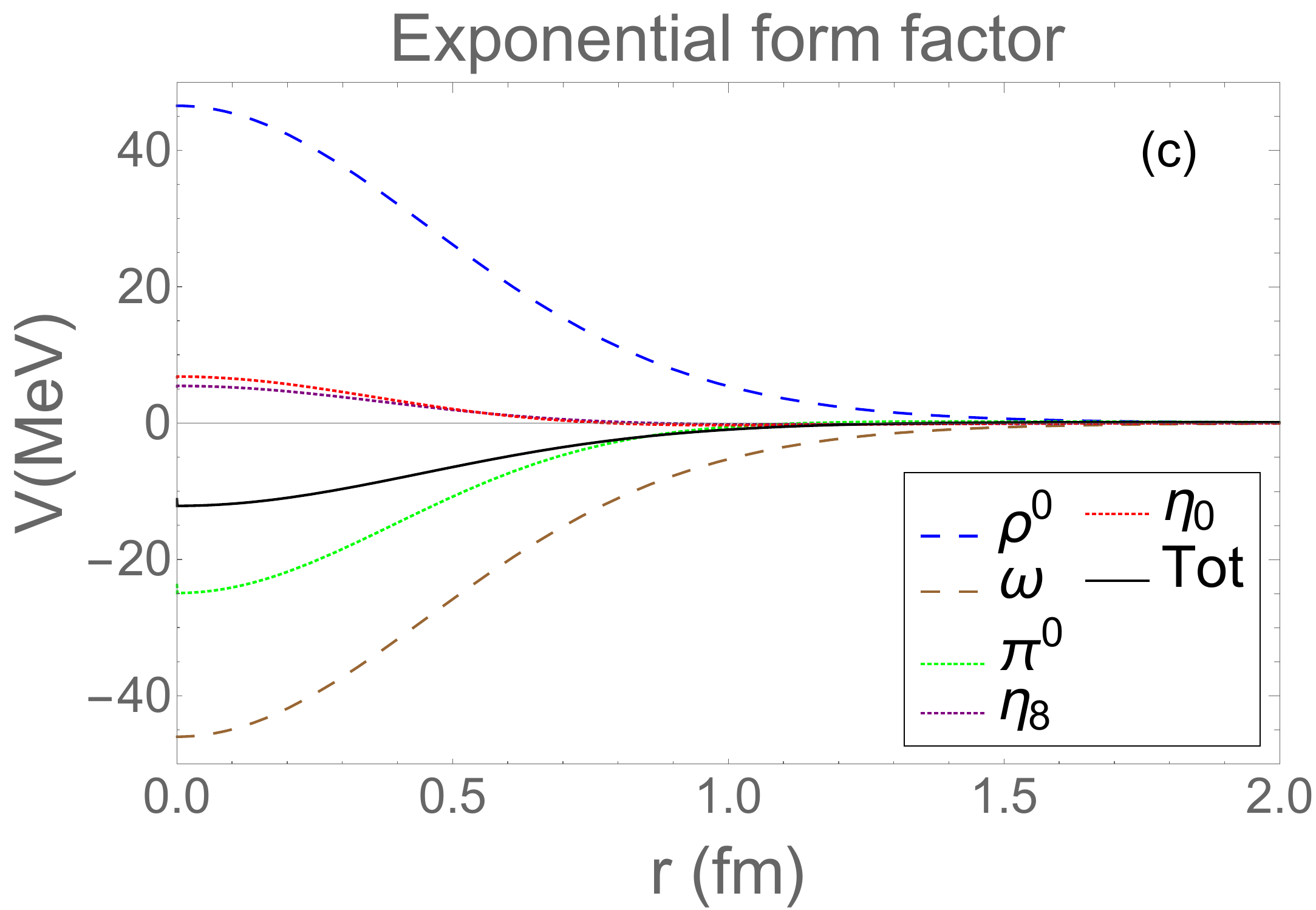}\includegraphics[width=0.3\linewidth]{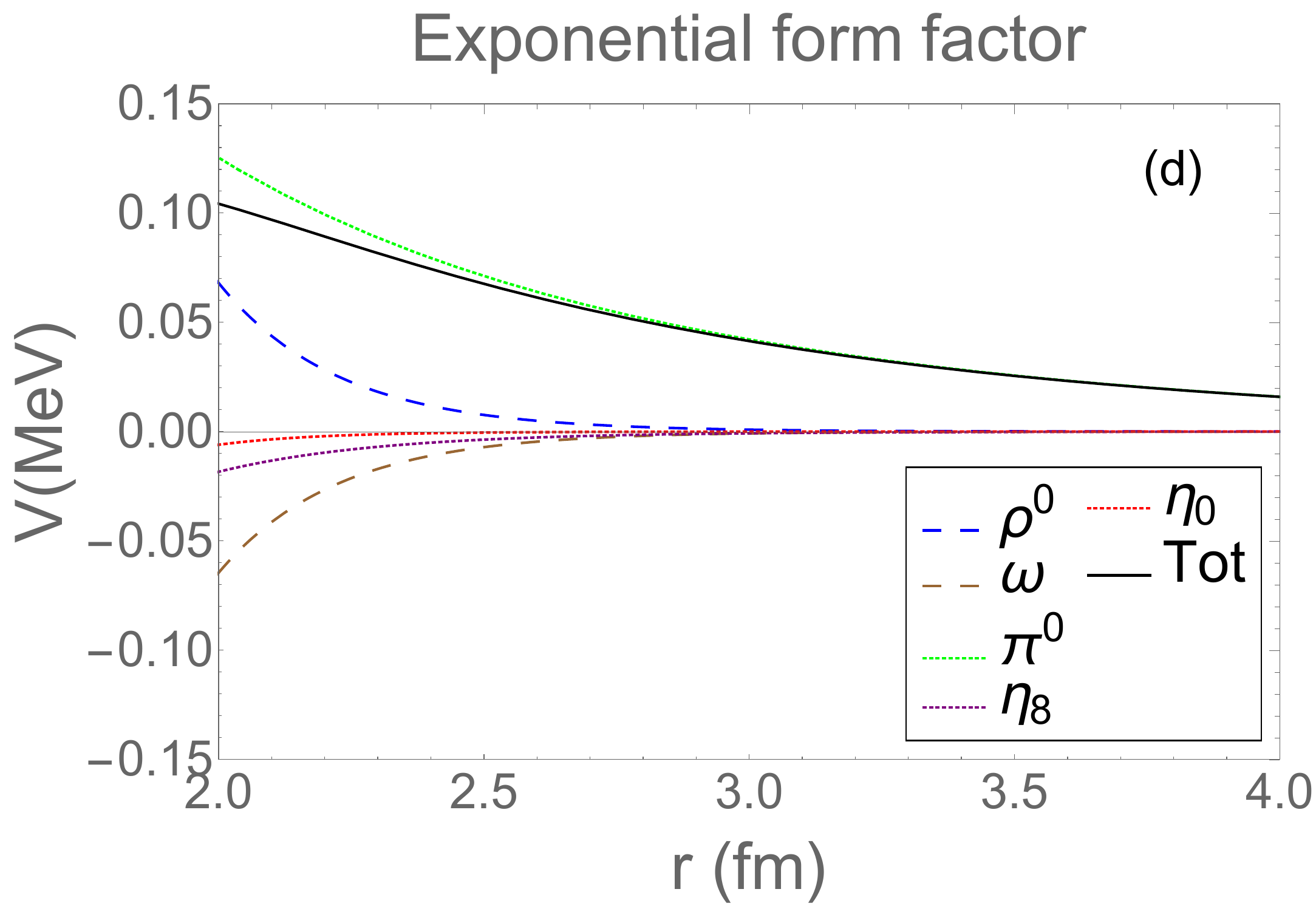}
    \caption{(Color online) The effective potential for S-wave $B^*\bar{B}^{*}$ to S-wave $B^*\bar{B}^{*}$ with different form factors.}
    \label{fig2:effective potential in different form factor}
\end{figure*}

As we mentioned above, in this work, we use the exponentially parameterized form factor in our calculation. For comparison, we take the $B^*\bar{B}^*\to B^*\bar{B}^*$ process as an example, and plot in Fig. \ref{fig2:effective potential in different form factor} the potentials with both the exponential form factor and the monopole one. We notice that
\begin{itemize}
    \item $V^{B^*\Bar{B}^*\to B^*\Bar{B}^*}_{\rho^0}$ and $V^{B^*\Bar{B}^*\to B^*\Bar{B}^*}_{\omega}$ have almost the same absolute value, but different signs, i.e., the contribution of $\rho-$ and $\omega-$exchange is approximately zero. So $V^{B^*\Bar{B}^*\to B^*\Bar{B}^*}$ is mainly contributed by $\pi-$, $\eta_8-$ and $\eta_0-$exchange;
\end{itemize}
\begin{itemize}
    \item the monopole form factor suppresses the contribution from the exchanged particle with heavier mass, which can be clearly seen from the situation of $\eta_0-$ exchange, since $m_{\eta_0}\simeq 0.96$ GeV;
\end{itemize} 
\begin{itemize}
    \item in the long interaction range, the pion-exchange subpotential is dominant with both monopole and exponential form factors; 
\end{itemize}
\begin{itemize}
    \item in the short and medium range, the vector-exchange subpotentials are compatibly larger than that of the pion-exchange with exponential form factor. However, for the monopole form factor, the pion-exchange contribution is still dominant.
\end{itemize}
The values of the masses of diquarks taken from Ref. \cite{Ferretti:2019zyh} are listed in Table \ref{tab3}, and the meson-masses are taken from PDG. Then we solve the Schr\"{o}dinger equation, and the numerical results are presented in Table \ref{tab4}. For $B\Bar{B}^*/B^*\Bar{B}^*/S\Bar{A}/A\Bar{A}$ and $B^*\Bar{B}^*/S\Bar{A}/A\Bar{A}$ systems, loosely bound states exist when the cut-off is reasonably chosen as $\Lambda\sim 1$ GeV. If the value of $\Lambda$ increases, the binding energies increase as well while the root-mean-square radii decrease. For both of these two systems, the D-wave contribution is much smaller than that of the S-wave. Besides, the meson-meson component is much larger than the diquark-antidiquark component. That is to say, both $Z_b(10610)$ and $Z_b(10650)$ can be explained as hadronic molecules mixing with a little diquark-antidiquark components.
\section{Summary}\label{summary}
In this work, we extend the $[U(3)_L\otimes U(3)_R]_{global}\otimes [U(3)]_{local}$ symmetry to the diquark sector and construct the Lagrangians containing diquarks. In this way, we can introduce the diquark-exchange interactions and study the nature of the $Z_b$ states as the mixture of hadronic molecule and diquark-antidiquark state. 
\begin{table*}[htb]
    \centering
    \caption{The obtained bound state solutions (binding energy E and root-mean-square radius $r_{\mathrm{RMS}}$) for the charged $Z_b$ system. $P_M$ is the probability of the channels $B\Bar{B}^*/B^*\Bar{B}(^3S_1), B\Bar{B}^*/B^*\Bar{B}(^3D_1), B^*\Bar{B}^*(^3S_1), B^*\Bar{B}^*(^3D_1)$ for $Z_b(10610)$, and $B^*\Bar{B}^*(^3S_1), B^*\Bar{B}^*(^3D_1)$ for $Z_b(10650)$, respectively. $P_D$ is the probability of the channels $S\Bar{A}/A\Bar{S}(^3S_1)$, $S\Bar{A}/A\Bar{S}(^3D_1)$, $A\Bar{A}(^3S_1)$, $A\Bar{A}(^3D_1)$.}
    \label{tab4}
    \begin{tabular}{ccccccc}
\toprule[1pt]  
State &\qquad $\Lambda \mathrm{(GeV)}$ & \qquad$E(\mathrm{MeV})$ &\qquad $r_{\mathrm{RMS}}(\mathrm{fm})$ & \qquad $P_{M}(\%)$ & \qquad $P_{D}(\%)$\\
\midrule
$Z_b(10610)$ &\qquad 1.0 &\qquad -1.83 &\qquad 1.64  &\qquad  94.24/0.09/2.85/0.02 &\qquad 2.18/0.04/0.57/0.01\\
             &\qquad 1.1 &\qquad -9.96 &\qquad 0.82  &\qquad  84.24/0.06/8.30/0.01 &\qquad 5.15/0.10/2.14/0.01\\
             &\qquad 1.2 &\qquad -26.14 &\qquad 0.57 &\qquad  73.78/0.15/13.55/0.00 &\qquad 7.83/0.16/4.53/0.01\\
$Z_b(10650)$ &\qquad 1.0 &\qquad -2.60 &\qquad 1.45  &\qquad  96.06/0.31 &\qquad 0.03/0.01/3.57/0.02\\
             &\qquad 1.1 &\qquad -9.01 &\qquad 0.88  &\qquad  92.34/0.28 &\qquad 0.09/0.02/7.23/0.04\\
             &\qquad 1.2 &\qquad -19.82 &\qquad 0.66 &\qquad  88.30/0.26 &\qquad 0.18//0.03/11.17/0.06\\
\bottomrule[1pt]
\end{tabular}
\end{table*}
In order to make the small distance interaction suppressed, we introduce the exponential form factor instead of the monopole form factor for each vertex, and calculate the effective potentials in the coordinate space analytically. We see that in the short and medium range, $\rho-$ or $\ omega-$ exchange contribution is larger than that of pion-exchange, while the pion-exchange is dominant in the long range. Besides, $B^{(*)-}\Bar{B}^{(*)0}\to B^{(*)-}\Bar{B}^{(*)0}$ processes are OZI suppressed which can be easily seen on the quark level. This is well depicted under the $[U(3)_L\otimes U(3)_R]_{global}\otimes [U(3)]_{local}$ symmetry, since the $\pi-$, $\eta_8-$, $\eta_0-$, $\rho-$ and $\omega-$exchange contributions cancel out. If taking the real masses of the exchanged particles which are different, the potential is non-zero corresponding to OZI violation.

Taking into account S- and D-wave contributions, we solve the coupled channel Schr\"{o}dinger equation which obeys the $U(1)\otimes U(1)\otimes ... \otimes U(1)$ symmetry. The numerical results show that both $Z_b(10610)$ and $Z_b(10650)$ can be explained as hadronic molecules slightly mixing with diquark-antidiquark states. If the cut-off is typically taken as $\Lambda=1$ GeV for both the $Z_b(10610)$ and the $Z_b(10650)$ states, the molecular components are about 97\% compared to around 3\% diquark-antidiquark state components. Besides, the S-wave contribution is much larger than that of the D-wave. Our work can help to understand the nature of the $Z_b$ states.
\section*{Acknowledgments}
We would like to thank Zhi-Peng Wang for very useful discussion. This project is supported by the Fundamental Research Funds for the Central Universities under Grant No. lzujbky-2022-sp02, the National Natural Science Foundation of China (NSFC) under Grants No. 11965016, 11705069 and 12247101, and the National Key Research and Development Program of China under Contract No. 2020YFA0406400.

\section*{Appendix: Calculation of The Functions $Y(\Lambda,m,r)$ and $U(\Lambda,m,r)$}
The definition of $Y(\Lambda,m,r)$ is shown in Eq. \eqref{eq52}, i.e.,
\begin{eqnarray}
Y(\Lambda,m,r)&=&\int \frac{d^3q}{(2\pi)^3}e^{i\vec{q}\cdot \vec{r}}\frac{1}{\vec{q}^2+m^2-i\epsilon}e^{2(q_0^2-\vec{q}^2)/\Lambda^2}.\nonumber\\
\end{eqnarray}
After integrating out the the azimuth and polar angle, we have
\begin{eqnarray}
Y(\Lambda,m,r)&=&-\frac{e^{2q_0^2/\Lambda^2}}{(2\pi^2r)}\frac{\partial}{\partial r}F(r,2/\Lambda^2)
\end{eqnarray}
where
\begin{eqnarray}
F(r,2/\Lambda^2)&=&\int_{-\infty}^\infty dq \frac{1}{{q}^2+m^2-i\epsilon}e^{-2q^2/\Lambda^2}e^{iqr}.
\end{eqnarray}
One can proof that the function $F(r,2/\Lambda^2)$ satisfies the following first-order differential equation
\begin{eqnarray}
&&\frac{\partial}{\partial \alpha}F(r,\alpha)-m^2F(r,\alpha)=-\int_{-\infty}^{\infty}dqe^{-q^2\alpha}e^{iqr}\nonumber\\
&=&-\sqrt{\frac{\pi}{\alpha}}e^{-\frac{r^2}{4\alpha}}\label{eq70}
\end{eqnarray}
with $\alpha=2/\Lambda^2$. Besides, we calculate $F(r,0)$ by using the residue theorem, i.e.,
\begin{eqnarray}
F(r,0)=\frac{\pi}{m}e^{-mr},
\end{eqnarray}
which can be treated as the initial condition of the differential equation \eqref{eq70}. Through solving this initial value problem, we have
\begin{eqnarray}
Y(\Lambda,m,r)&=&-\frac{e^{2q_{0}^{2}/\Lambda^2}}{(2\pi)^2r}\frac{\partial}{\partial r} \Bigg\{e^{2m^{2}/\Lambda^2}\frac{\pi}{2m}\Bigg[e^{mr}\nonumber\\
&&+e^{-mr}-e^{mr}\text{erf}\left(\frac{\Lambda r}{2\sqrt{2}}+\frac{\sqrt{2}m}{\Lambda}\right)\nonumber\\
&&+e^{-mr}\text{erf}\left(\frac{\Lambda r}{2\sqrt{2}}-\frac{\sqrt{2}m}{\Lambda}\right)\Bigg]\Bigg\},\label{eq9}
\end{eqnarray}

Using the same method, as a product, we also calculate the following function
\begin{eqnarray}
U(\Lambda,m,r)&=&\int \frac{d^3q}{(2\pi)^3}e^{i\vec{q}\cdot \vec{r}}\frac{1}{\vec{q}^2-m^2-i\epsilon}e^{2(q_0^2-\vec{q}^2)/\Lambda^2}\nonumber\\
&=&\frac{e^{2q_0^2/\Lambda^2}}{(2\pi)^2r}\frac{\partial}{\partial r}\left\{\pi\left[-\frac{1}{2im}\left(e^{-imr}\text{erf}\left(\frac{r\Lambda}{2\sqrt{2}}\right.\right.\right.\right.\nonumber\\
&&\left.\left.-\frac{\sqrt{2}im}{\Lambda}\right)-e^{imr}\text{erf}\left(\frac{r\Lambda}{2\sqrt{2}}+\frac{\sqrt{2}im}{\Lambda}\right)\right)\nonumber\\
&&\left.\left.-\frac{i}{m}\cos(mr)\right]e^{-2m^2/\Lambda^2}\right\}.\label{eq10}
\end{eqnarray}


\end{document}